\documentclass{emulateapj}
\input psfig.sty

\newcommand{\beq}{\begin{equation}}
\newcommand{\eeq}{\end{equation}}

\begin{document}

\title{Measuring the Three-Dimensional Structure of Galaxy Clusters.
I. Application to a Sample of 25 Clusters.}
\shorttitle{Measurements of the 3-D Structure of Galaxy Clusters}

\author{Elisabetta De Filippis\altaffilmark{1}}
\author{Mauro Sereno\altaffilmark{2,3,4}}
\author{Mark~W.Bautz\altaffilmark{1}}
\author{Giuseppe Longo\altaffilmark{3,4}}

\altaffiltext{1}{Center for Space Research,
               Massachusetts Institute of Technology,
               70 Vassar Street, Building 37,
               Cambridge, MA 02139; bdf@space.mit.edu, mwb@space.mit.edu}

\altaffiltext{2}{INAF-Osservatorio Astronomico di Capodimonte,
                Salita Moiariello, 16
                80131 Naples, Italy}

\altaffiltext{3}{Dipartimento di Scienze Fisiche, Universit\`{a} degli Studi di Napoli ``Federico II'',
                Via Cinthia, Compl. Univ. Monte S. Angelo,
                80126 Naples, Italy}

\altaffiltext{4}{INFN-Sez. Napoli, Compl. Univ. Monte S. Angelo,
                80126 Naples, Italy; sereno@na.infn.it; longo@na.infn.it}

\shortauthors{De Filippis et al.}

\begin{abstract}
We discuss a method to constrain the intrinsic shapes of galaxy clusters
by combining X-Ray and Sunyaev-Zeldovich observations. 
The method is applied to a sample of $25$ X-Ray selected clusters, 
with measured Sunyaev-Zeldovich temperature
decrements. The sample turns out to be slightly biased, with 
strongly elongated clusters preferentially aligned along the line of 
sight. This result demonstrates that X-Ray selected cluster 
samples may be affected by morphological and orientation effects even if
a relatively high threshold signal-to-noise ratio is used to select the 
sample.  A large majority of the clusters in our sample exhibit a marked
triaxial structure; the spherical hypothesis is strongly rejected 
for most sample members. 
Cooling flow clusters do not show preferentially
regular morphologies.
We also show that  identification of multiple gravitationally-lensed 
images, together with measurements of the Sunyaev-Zeldovich effect and
X-Ray surface brightness, can provide a simultaneous determination of
the three-dimensional structure of a cluster, of the Hubble constant,
and the cosmological energy density parameters.

\end{abstract}


\section{Introduction}
The intrinsic, three-dimensional (hereafter 3-D) shape of clusters 
of galaxies is an important cosmological probe. The structure of 
galaxy clusters is sensitive to the mass density in the universe, so 
knowledge of this structure can help in discriminating between different 
cosmological models. It has long been clear that the
formation epoch of galaxy clusters strongly depends on the matter
density parameter of the universe \citep{Ric92}. The growth of
structure in a high-matter-density universe is expected to continue
to the present day, whereas in a low density universe the fraction of
recently formed clusters, which are more likely to have substructure, 
is lower. Therefore, a sub-critical value of the density parameter $\Omega_{\rm
M0}$ favors clusters with steeper density profiles and rounder
isodensity contours.  Less dramatically, a
cosmological constant also delays the formation epoch of clusters, 
favoring the presence of structural irregularity~\citep{Suw03}.\\
An accurate knowledge of intrinsic cluster shape is also required 
to constrain structure formation models via observations of 
clusters. The asphericity of dark halos affects the 
inferred central mass density of clusters, the 
predicted frequency of gravitational arcs, 
nonlinear clustering (especially high-order clustering
statistics) and dynamics of galactic satellites (see~\cite{Jin02} and
references therein).\\
Asphericity in the gas density distribution of clusters of galaxies is
crucial in modeling X-Ray morphologies and in using clusters as 
cosmological tools.
\citep{Ina95,Coo98,Sul99}. Assumed  cluster shape
strongly affects absolute distances obtained from X-Ray/Sunyaev-Zeldovich (SZ) measurements, as well as relative distances obtained from baryon fraction constraints
~\citep{All04,Coo98}.  Finally, all cluster mass measurements
derived from X-Ray and dynamical observations are sensitive to the 
assumptions about cluster symmetry.\\
Of course, only the two-dimensional (2-D) projected properties of
clusters  can be observed. The question of how to deproject
observed images is a well-posed inversion problem that has been studied 
by many authors~\citep{Luc74,Ryd96,Reb00}. 
Since  information is lost in the process of projection  it is 
in general impossible to derive the  intrinsic 3-D 
shape of an astronomical object from a single observation.
To some extent, however, one can overcome this degeneracy by combining 
observations in different wavelengths.  
For example, \cite{Zar98,Zar01} introduced a model-independent
method of image deprojection. This inversion method uses X-Ray,
radio and  weak lensing maps to infer
the underlying 3-D structure for an axially symmetric distribution. 
\cite{Reb00} proposed a parameter-free algorithm for the 
deprojection of observed two dimensional cluster images, again using weak
lensing, X-Ray surface brightness and SZ imaging. 
The 3-D gravitational potential was assumed to be axially symmetric and
the inclination angle was required as an input parameter. Strategies for
determining the orientation have been also discussed. 
\cite{Dor01} proposed a method that, with a perturbative approach and 
with the aid of SZ and weak lensing data, could predict the cluster 
X-Ray emissivity without resolving the full 3-D structure of the cluster.
The degeneracy between the distance to galaxy clusters and
the elongation of the cluster along the line of sight (l.o.s.) was thoroughly
discussed by~\cite{Fox02}. They 
introduced a specific method for finding the intrinsic 3-D shape of 
triaxial cluster and, at the same time, measuring the 
distance to the cluster corrected for asphericity, so providing an
unbiased estimate of the Hubble constant $H_0$. 
\cite{Lee04} recently proposed a theoretical method to reconstruct the shape 
of triaxial dark matter halos using X-Ray and SZ data. The Hubble constant and the
projection angle of one principal axis of the cluster on the plane of
the sky being independently known, they constructed a numerical
algorithm to determine the halo eccentricities and orientation.
However, neither~\cite{Fox02}
nor~\cite{Lee04} apply their method to real data.\\
In this paper we focus on X-Ray surface
brightness observations and SZ temperature decrement measurements. 
We show how the intrinsic 3-D shape of a cluster of galaxies 
can be determined through joint analyses of these data, given an 
assumed cosmology. We constrain the triaxial structure 
of a sample of observed clusters of galaxies with
measured X-Ray and SZ maps. To break the degeneracy between
shape and cosmology, we adopt cosmological parameters which have 
been relatively well-determined from  measurements of the 
cosmic microwave background (CMB) anisotropy,  Type Ia 
supernovae and the spatial distribution of galaxies. 
We also show how, if  multiply-imaging gravitational lens systems 
are observed, a joint analysis of strong lensing, X-Rays and SZ data allows a
determination of both the 3-D shape of a cluster and
the geometrical properties of the universe.\\
The paper is organized as follows. 
The basic dependencies of cluster X-Ray emission and the SZE on 
geometry are reviewed in \S~\ref{sec:multi_wave}. In
\S~\ref{sec:combin_datasets}, we show how to reconstruct the 
3-D cluster structure from these data, 
presuming cosmological parameters to be known. 
In passing we note how the addition of suitable strong gravitational
lensing data can constrain the cosmological parameters as well, although
we do not impose lensing constraints in this paper. 
We then turn to face the data.  Our cluster sample 
is introduced in \S~\ref{sec:data_samp}, and  in
\S~\ref{sec:morph_2D}, we present 2-D X-Ray surface brightness parameters 
for each sample member. 
The triaxial structure of the clusters is then estimated and analyzed in 
\S~\ref{sec:tria}. 
\S~\ref{sec:disc} is devoted to a summary and discussion of the results. 
In Appendix~\ref{sec:triaxial}, we
provide details on the triaxial ellipsoidal $\beta$-model, used to describe 
the intra-cluster gas distribution, while Appendix~\ref{sec:inclination}
is devoted to a discussion of the consequences of our assumption of 
clusters being triaxial ellipsoids aligned along the line of sight. 
In Appendix~\ref{sec:lensing} the identifications of multiple sets of images of background 
galaxies in strong lensing events is discussed.
Throughout this paper, unless otherwise stated, we quote errors at the $68.3\%$ confidence
level.

\section{Multi-Wavelength Approach}
\label{sec:multi_wave}
In this section, we summarize the relationships between SZ
and X-Ray observables, on the one hand, and cluster shape and distance 
on the other. 

\subsection{The Sunyaev-Zeldovich Effect}
The gravitational potential wells of galaxy clusters contain 
plasma at temperatures of about $k_{\rm B} T_{\rm e} \approx 8$-$10\ {\rm
keV}$. CMB  photons that pass through a
cluster interact with the energetic electrons of its hot intra-cluster medium (ICM) through
inverse Compton scattering, with a probability $\tau\sim 0.01$. This
interaction causes a small distortion in the CMB spectrum, known as
the Sunyaev-Zeldovich effect (SZE)~\citep{Sun70,Bir99}, which is
proportional to the electron pressure integrated along the l.o.s.,
i.e. to the first power of the plasma density. The measured
temperature decrement $\Delta T_{\rm SZ}$ of the CMB is given by:
\begin{equation}
\label{eq:sze1}
\frac{\Delta T_{\rm SZ}}{T_{\rm CMB}} = f(\nu, T_{\rm e}) \frac{ \sigma_{\rm T} k_{\rm B} }{m_{\rm e} c^2}
\int _{\rm l.o.s.}n_e T_{\rm e} dl \
\end{equation}
where $T_{\rm e}$ is the temperature of the ICM, $k_{\rm B}$ the
Boltzmann constant, $T_{\rm CMB} =2.728^{\circ}$K is the temperature of the
CMB, $\sigma_{\rm T}$ the Thompson cross section, $m_{\rm e}$ the
electron mass, $c$ the speed of light in vacuum and $f(\nu, T_{\rm
e})$ accounts for frequency shift and relativistic corrections.\\
If we assume that the ICM is described by an isothermal triaxial
$\beta$-model distribution, substituting Eq.~(\ref{eq:tri5}) into
(\ref{eq:sze1}) with $m=1$, we obtain:
\begin{equation}
\label{eq:sz2}
\Delta T_{\rm SZ} = \Delta T_0 \left( 1+ \frac{\theta_{1}^2+e_{\rm proj}^2 \theta_{2}^2}{\theta_{c,\rm proj}^2}
\right)^{1/2-3\beta/2}
\end{equation}
where $\Delta T_0$ is the central temperature decrement which includes
all the physical constants and the terms resulting from the l.o.s.
integration
\begin{eqnarray}
\label{eq:sze3}
\Delta T_0 &\equiv &T_{\rm CMB} f(\nu, T_{\rm e}) \frac{ \sigma_{\rm T} k_{\rm B} T_{\rm e}}{m_{\rm e} c^2}n_{e0} \sqrt{\pi}\nonumber \\
&\times&  \frac{D_{\rm c}\theta_{\rm c,proj}}{h^{3/4}}\sqrt{\frac{e_1 e_2}{e_{\rm proj}}} g\left(\beta/2\right)
\end{eqnarray}
with:
\begin{center}
$$g(\alpha)\equiv\frac{\Gamma \left[3\alpha-1/2\right]}{\Gamma \left[3 \alpha\right]}. $$
\end{center}
$D_{\rm c}$ is the angular diameter distance to the cluster, 
$\theta_i \equiv x_{i,\rm obs}/D_{\rm c}$ is the projected angular
position (on the plane of the sky) of the intrinsic orthogonal coordinate 
$x_{i,\rm obs}$, $h$ is a function of the cluster shape and 
orientation (Eq.~\ref{eq:tri3}), $e_{\rm proj}$ is the axial ratio 
of the major to the minor axes of the observed projected isophotes and 
$\theta_{c,\rm proj}$ the projection on the plane of the sky (p.o.s.) 
of the intrinsic angular core radius (Eq.~\ref{eq:tri6}).
In a Friedmann-Lema\^{\i}tre-Robertson-Walker universe filled with
pressure-less matter and with a  cosmological
constant, the angular diameter distance between an observer at a
redshift $z_{\rm d}$ and a source at $z_{\rm c}$ is:
\begin{equation}
\label{eq:crit3}
\left. D_{\rm c}\right|_{\rm Cosm}(z_{\rm d}, z_{\rm c})=\frac{c}{H_0}\frac{1}{1+z_{\rm
c}}\frac{1}{|\Omega_{\rm K0}|} {\rm Sinn} \left( \int_{z_{\rm
d}}^{z_{\rm c}} \frac{|\Omega_{\rm K0}|}{{\cal E}(z)} dz \right)
\end{equation}
with
\begin{eqnarray}
\label{eq:dist1}
{\cal E}(z) & \equiv & \frac{H(z)}{H_0} \\ & =& \sqrt{ \Omega_{\rm M0}
(1+z)^3+ \Omega_{\Lambda 0} + \Omega_{\rm K0}(1+z)^2}
\nonumber
\end{eqnarray}
where $H_0$, $\Omega_{\rm M0}$ and $\Omega_{\Lambda 0}$ are the Hubble
parameter, the normalized energy density of pressure-less matter and
the reduced cosmological constant at $z=0$, respectively. $\Omega_{\rm
K0}$ is given by $\Omega_{\rm K0}\equiv 1- \Omega_{M
0}-\Omega_{\Lambda 0}$, and Sinn is defined as being $\sinh$ when
$\Omega_{\rm K0}>0$, $\sin$ when $\Omega_{\rm K0}<0$, and as the
identity when $\Omega_{\rm K0}=0$. A more general expression of the
angular diameter distance, also accounting for dark energy and
inhomogeneity in matter distribution, can be found
in~\cite{Ser01,SerAl02}.

\subsection{X-Ray Surface Brightness}
Cluster X-Ray emission is due to bremsstrahlung and line radiation resulting 
from electron-ion collisions; the X-Ray surface brightness $S_X$ is proportional to
the integral along the l.o.s. of the square of the 
electron density:
\begin{equation}
S_X = \frac{1}{4 \pi (1+z_{\rm c})^4} \int _{\rm l.o.s.} n_e^2
\Lambda_{eH} dl
\label{eq:sxb0}
\end{equation}
where $\Lambda_{eH}$ is the X-Ray cooling function of the ICM in the
cluster rest frame. Substituting Eq.~(\ref{eq:tri5}) into
(\ref{eq:sxb0}) with $m=2$, we get:
\begin{equation}
S_X = S_{X0} \left( 1+ \frac{\theta_{1}^2+e_{\rm proj}^2
\theta_{2}^2}{\theta_{c,\rm proj}^2} \right)^{1/2-3 \beta}
\label{eq:sxb1}
\end{equation}
where the central surface brightness $S_{X0}$ reads:
\begin{equation}
\label{eq:sxb2}
S_{X0} \equiv \frac{ \Lambda_{eH}\ \mu_e/\mu_H}{4 \sqrt{\pi} (1+z_{\rm c})^4}
n_{e0}^2  \frac{D_{\rm c}
\theta_{\rm c,proj}}{h^{3/4}}\sqrt{\frac{e_1 e_2}{e_{\rm proj}}}\ g(\beta)
\end{equation}
$\mu$ is the molecular weight given by: $\mu_i\equiv \rho/n_im_p$.

\section{Combining Heterogeneous Data Sets}
\label{sec:combin_datasets}
Here we discuss how  2-D SZE and X-Ray maps of a cluster
can be used to constrain its 3-D shape.
We follow a parametric approach. We model the cluster 
using an isothermal, triaxial
$\beta$ profile, and adopt a concordance model for the cosmological 
distance-redshift relationships. Details of the cluster model are given
in Appendix~\ref{sec:triaxial}.\\
This model has long been used to describe
the electron distributions of galaxy clusters. It was originally
introduced specifically for dynamically relaxed, isothermal
clusters~\citep{Cav78}, but it was then observed to fit 
the X-Ray emission of most galaxy clusters reasonably
well. A serious drawback of this model is
that electron density profiles with extreme axial ratios lead either to
unlikely total mass density distributions, i.e. dumbbell shaped
clusters, or to regions with unphysical (negative) density. Nevertheless,
we believe its extreme versatility makes it a useful tool for our 
purposes.\\
For an ellipsoidal distribution, the 3-D shape of a
cluster is described by two axis ratios, $e_1$ and $e_2$, and the
orientation of the cluster is fixed by three Euler angles,
$\theta_{\rm Eu}, \phi_{\rm Eu}$ and $\psi_{\rm Eu}$. As shown
in Eqs.~(\ref{eq:tri0},~\ref{eq:gl2}), in our model the density profile of a
cluster is characterized by three additional parameters: the central
density $n_{e0}$, the slope $\beta$ and a core radius, $r_{\rm c3}$.
Under the hypothesis of isothermal ICM, a single value, $T_{\rm e}$,
characterizes the temperature profile of the cluster. In all, nine
parameters describe the cluster.\\
As discussed in the previous section, the cosmological dependence of the 
model enters through the luminosity-redshift relationship.
For a flat model  universe, this relationship is in turn 
determined by two parameters: the Hubble
constant, $H_0$, and the matter density $\Omega_{\rm M0}$.\\
A projected axis ratio, $e_{\rm proj}$, and an orientation angle,
$\psi$, characterize a family of ellipses in the p.o.s. derived from
the 2-D projection of 3-D ellipsoids. By fitting
an elliptical profile to the X-Ray and/or SZE data, these two
parameters can (in principle) be measured. Two other observables, 
the slope $\beta$ of the profile and the projected core 
radius $\theta_{c,\rm proj}$ can also be determined from data. 
Two independent geometrical constraints relate 2-D and 3-D
quantities (Eqs.~\ref{eq:tri4e},~\ref{eq:tri4f}).\\
So far  we have discussed only quantities derivable from spatial distributions.
Besides these, the cluster central electron density and the temperature of the 
ICM can be inferred from X-Ray 
observations with sufficient energy resolution. The observed
values of the central temperature decrement, $\Delta T_0$ in
Eq.~(\ref{eq:sze3}) and of the central surface brightness, $S_{X0}$ in
Eq.~(\ref{eq:sxb2}), provide two further constraints. 
If some assumption is made on the orientation of the cluster, with eight 
independent equations and eight unknown physical parameters 
a full determination of the cluster shape can be obtained.
If a rotational ellipsoidal morphology is chosen, a lower number of parameters 
is needed to describe the three-dimensional shape of clusters; in this case 
no additional assumption on the inclination is required allowing a full 
determination of the cluster shape and orientation.
This case is treated in details in a subsequent paper~\citep{Ser04}.

\subsection{Adding Strong Lensing Data}
We wish to point that, even though we do not do so in this work,  
strong lensing data can be combined with the X-Ray and SZE observations
to break  the degeneracy between the intrinsic shape
of the lensing cluster and the cosmological parameters.  If this were done,
one could obtain simultaneous constraints on the cluster parameters and 
on the cosmology.\\
In particular, each set of strong gravitational images 
identified in a cluster  strongly constrains the mass distribution of the
lens.  The convergence $k$ depends on the cosmology only
through the ratio of distances $D_{\rm cs}/D_{\rm s}$. Therefore $k$ depends
only the cosmological density parameters $\Omega_i$, and not on the
Hubble constant $H_0$. The value of $k_0$ changes according to the
redshift $z_s$ of the lensed source. Image systems produced by 
sources at different redshifts probe independent values of the ratio $D_{\rm
cs}/{D_{\rm s}}$.\\
Each image system provides a constraint on 
central value of the convergence $k_0(z_{\rm s})$ 
(Eq.~\ref{eq:gl6b}). In turn,
if both the positions and the redshift of a multiple image system are
known, each measured value of $k_0$ provides, through
Eq.~(\ref{eq:gl6b}), a further independent constraint on the
cosmological energy densities. Each  multiply-imaged source
provides an independent constraint which
relates the cosmological parameters $\Omega_i$ to the 3-D
shape and orientation of the cluster ($h^{3/4}/\sqrt{e_1 e_2}$).
With a sufficient number of image systems, then, 
a measure of both the intrinsic shape and orientation of the cluster 
and a simultaneous estimate of all cosmological parameters involved 
can therefore be performed.

\subsection{Angular Diameter Distances from  X-Ray
and SZE Observations for Triaxial Clusters}
It is of course well known that the angular diameter
distance to a spherically symmetric cluster can be be inferred 
from microwave decrement and X-Ray data. The angular diameter 
distance enters the SZE and the X-Ray emission through
a characteristic length-scale of the cluster along the l.o.s.
SZE and X-Ray emission depend differently on the density of ICM, and
therefore also on the assumed cosmology. A joint analysis of SZE
measurements and X-Ray imaging observations, together with the 
assumption of spherical symmetry, thus can 
yield the distance to the cluster \citep{Bir99,Ree02}.  Specifically, 
one can solve Eqs.~(\ref{eq:sze3}) and (\ref{eq:sxb2}) for the angular
diameter distance $D_{\rm c}$, by eliminating $n_{\rm e0}$.\\
More generally, for a triaxial cluster  the inferred angular diameter distance
takes the form:
\begin{eqnarray}
D_{\rm c} & =&  \left. D_{\rm c}\right|_{\rm Exp}
\frac{\theta_{\rm c,proj}}{\theta_{\rm c3}}h^{1/2} \nonumber \\ &  = & \left. D_{\rm
c}\right|_{\rm Exp} h^{3/4} \left(\frac{e_{\rm proj} }{e_1 e_2}
\right)^{1/2}
\label{eq:dis2}
\end{eqnarray}
where $\left. D_{\rm c}\right|_{\rm Exp}$ is an experimental quantity given by:
\begin{eqnarray}
\label{eq:obl7}
\left. D_{\rm c}\right|_{\rm Exp} &= &\frac{\Delta T_0^2}{S_{\rm X0}}
\left( \frac{m_{\rm e} c^2}{k_{\rm B} T_{e0} } \right)^2 \frac{g\left(\beta\right)}{g(\beta/2)^2\ \theta_{\rm c,proj}}\nonumber \\
& \times &\frac{\Lambda_{eH0}\ \mu_e/\mu_H}{4 \pi^{3/2}f(\nu,T_{\rm e})^2\ T^2_{\rm CMB}\ \sigma_{\rm T}^2\ (1+z_{\rm c})^4}.
\end{eqnarray}
Under the assumption of spherical symmetry, the 3-D morphology of the cluster is
completely known: $h=e_1=e_2=1$, $\theta_{\rm Eu}=\varphi_{\rm Eu}=\psi_{\rm Eu}=0$ and
the observed major core radius $\theta_{\rm c, proj}$ reduces to $\theta_{\rm c}$. 
Hence Eq.~(\ref{eq:dis2}) becomes:
$$D_{\rm c} =\left. D_{\rm c}\right|_{\rm Exp}$$
and the cluster angular diameter
distance can therefore be obtained directly from Eq.~(\ref{eq:obl7}).
The standard approach in the past decades has been to take advantage of this 
possibility to estimate $D_{\rm c}$ under the assumption of spherical symmetry, 
in order to constrain the underlying cosmology. 
Since in fact it is also true that $D_{\rm c}=\left. D_{\rm c}\right|_{\rm Cosm}$, 
through Eq.~(\ref{eq:crit3}) an estimate of $H_0$ can 
be obtained if $\Omega_{M0}$ and $\Omega_{\Lambda 0}$ are
known from independent observations~\citep{Bir99,Ree02,Mas01}.\\
The same approach clearly cannot be applied when the assumption of spherical 
symmetry is relaxed and clusters are considered as more general triaxial systems. In this case an estimate of the axis ratios, shape and orientation parameters 
is required before $D_{\rm c}$ can be computed.
Conversely, if $D_{\rm c}$ is known from the redshift and prior 
knowledge of the cosmology, then the X-Ray and SZE data 
can be used to constrain the 3-D morphology
of the cluster.  In this paper we will follow this latter approach. 
We assume the values of $\Omega_i$ and of $H_0$ to be known; 
$\left. D_{\rm c}\right|_{\rm Cosm}$ can be then determined through 
Eq.~(\ref{eq:crit3}). We will then use
Eq.~(\ref{eq:dis2}) to infer the 3-D morphology 
of a sample of galaxy clusters.
We believe the cosmological distance scale is now known with sufficient
accuracy to warrant our approach. An impressive body of evidence from 
CMB anisotropy, Type Ia supernovae, galaxy clustering, large-scale structure,
and the Ly$\alpha$ forest ~\citep{Wan00} are consistent 
with a the picture of a universe with
sub-critical cold dark matter energy density and with two-thirds
of the critical density being in the form of dark energy. \cite{Teg04}
combine the three dimensional power spectrum from over $200,000$
galaxies in the Sloan Digital Sky Survey with the first-year Wilkinson
Microwave Anisotropy Probe (WMAP) data~\citep{Spe03} to measure
cosmological parameters. Their results are consistent with a flat
($\Omega_{\rm K}= 0$) cosmological model with
$H_0=70^{+4}_{-3}\ {\rm km\ s^{-1}\ Mpc^{-1}}$, $\Omega_{\rm M}=0.30 {\pm} 0.04$ 
and with a non-zero cosmological constant. 
Thanks to the high precision to which cosmological parameters are known, 
we are able to constrain measurements of $\left. D_{\rm c}\right|_{\rm Cosm}$ 
for the sample objects within a $5\%$ error.

\subsection{Cluster Elongation Along the Line of Sight}
\label{sec:application}
For the remainder of this paper, we will assume that every cluster
is triaxial, with one principal axis aligned along the l.o.s. 
(see \S~\ref{sec:combin_datasets}). 
In Appendix~\ref{sec:inclination} we show that the magnitude of the systematic 
error in inferred elongation parameters cause by such assumption is small 
compared to the uncertainties arising from the observational errors.
Such a straightforward assumption also 
leads to an extremely simple formalism to describe 
the resulting three-dimensional shape of clusters, reducing 
the errors caused by the uncertainties in the observational data.

The assumption that the cluster is aligned along the l.o.s. implies:
$\theta_{\rm Eu}=\varphi_{\rm Eu}=\psi_{\rm Eu}=0,\ h=1,\ j=e_1^2,\
k=0,\ {\rm and}\ l=e_2^2$; (see Appendix~\ref{sec:triaxial}).  
We label axes so that major axis is parallel to the 
$x_1$; then the projected axial ratio and core
radius are: $e_{\rm proj}=v_2/v_1$ and $\theta_{c,\rm
proj}=\theta_{\rm c}/v_1$. The angular diameter distance becomes:
\begin{equation}
\label{eq:simp1}
D_{\rm c} = \left. D_{\rm c}\right|_{\rm Cosm} = \left. D_{\rm c}\right|_{\rm Exp} \frac{v_3}{v_1}.
\end{equation}
We now introduce the elongation $e_{\rm l.o.s.}$, defined as the ratio of the
radius of the cluster along the l.o.s. to its major axis in the p.o.s.,
\begin{eqnarray}
e_{\rm l.o.s.}  & \equiv & \frac{v_1}{v_3}  \\ &
= &\frac{\left. D_{\rm c}\right|_{\rm Exp}}{\left. D_{\rm c}\right|_{\rm Cosm} (z,H_0,\Omega_M)}.
\label{eq:simp3}
\end{eqnarray}
Spherical clusters have the same radius along the l.o.s. and in
the p.o.s. and for them $e_{\rm l.o.s.} = 1$. Clusters
which are instead more or less elongated along the l.o.s.
than in  the p.o.s. will have values of $e_{\rm l.o.s.} >
1$ or $e_{\rm l.o.s.} < 1$, respectively.

\section{Data sample}
\label{sec:data_samp}
We now apply the formalism 
described in \S~\ref{sec:application} to a sample of galaxy clusters 
to infer new information about the extent of the clusters along the l.o.s. \\
We use two samples of clusters for which combined X-Ray and SZ
analysis has already been reported. The sample discussed by~\cite{Ree02} 
consists of $18$ X-Ray selected clusters with $z\geq 0.14$ and $\delta\geq
-15\degr$ and $L_{\rm X}(0.1 - 2.4\ {\rm keV})\geq 5 {\times} 10^{44}
h_{50}^{-2}\ {\rm erg\ s}^{-1}$ and for which high S/N detections of
SZE, high-S/N X-Ray imaging and electron temperatures were available.
To these we add the sample of~\cite{Mas01}, which  contains $7$ clusters 
from X-Ray-flux-limited sample of~\cite{Ebe96}. Details
on the completeness of the latter subsample are given by~\cite{Mas00}.\\
Basic  data for our $25$ clusters, including previously published 
redshift, plasma temperature and microwave decrement 
information~\citep{Ree02,Mas01} are presented in Table~\ref{tab:sample}.

\tabletypesize{\scriptsize}
\def\arraystretch{1.0}
\begin{deluxetable}{lrrr}
\tablecolumns{7}
\tablewidth{0pt}
\tablecaption{Clusters in the Sample}
\tablehead{
\colhead{Cluster}  &\colhead{$z$}  &\colhead{$k_{\rm B}T_{\rm e}$}&\colhead{$\Delta T_0$}\\
\colhead{}         &\colhead{}   & \colhead{(keV)}& \colhead{($\mu K$)}}
\startdata
MS 1137.5+6625  &$0.784$  &$5.7^{+1.3}_{-0.7}$ 	    &$-818^{+98}_{-113}$\\ 
MS 0451.6-0305  &$0.550$  &$10.4^{+1.0}_{-0.8}$     &$-1431^{+98}_{-105}$\\ 
Cl 0016+1609    &$0.546$  &$7.55^{+0.72}_{-0.58}$   &$-1242^{+105}_{-105}$\\ 
RXJ1347.5-1145	&$0.451$  &$9.3^{+0.7}_{-0.6}$      &$-3950^{+350}_{-350}$\\ 
A 370 		&$0.374$  &$6.6^{+0.7}_{-0.5}$      &$-785^{+118}_{-118}$\\ 
MS 1358.4+6245	&$0.327$  &$7.48^{+0.50}_{-0.42}$   &$-784^{+90}_{-90}$\\ 
A 1995		&$0.322$  &$8.59^{+0.86}_{-0.67}$   &$-1023^{+83}_{-77}$\\ 
A 611		&$0.288$  &$6.6^{+0.6}_{-0.6}$      &$-853^{+120}_{-140}$\\ 
A 697 		&$0.282$  &$9.8^{+0.7}_{-0.7}$      &$-1410^{+160}_{-180}$\\ 
A 1835 		&$0.252$  &$8.21^{+0.19}_{-0.17}$   &$-2502^{+150}_{-175}$\\
A 2261 		&$0.224$  &$8.82^{+0.37}_{-0.32}$   &$-1697^{+200}_{-200}$\\ 
A 773 		&$0.216$  &$9.29^{+0.41}_{-0.36}$   &$-1260^{+160}_{-160}$\\ 
A 2163 		&$0.202$  &$12.2^{+1.1}_{-0.7}$     &$-1900^{+140}_{-140}$\\ 
A 520 		&$0.202$  &$8.33^{+0.46}_{-0.40}$   &$-662^{+95}_{-95}$\\ 
A 1689 		&$0.183$  &$9.66^{+0.22}_{-0.20}$   &$-1729^{+105}_{-120}$\\ 
A 665 		&$0.182$  &$9.03^{+0.35}_{-0.31}$   &$-728^{+150}_{-150}$\\ 
A 2218 		&$0.171$  &$7.05^{+0.22}_{-0.21}$   &$-731^{+125}_{-100}$\\ 
A 1413 		&$0.142$  &$7.54^{+0.17}_{-0.16}$   &$-856^{+110}_{-110}$\\ 
A 2142 		&$0.091$  &$7.0\pm0.2$ 		    &$-437^{+25}_{-25}$\\ 
A 478 		&$0.088$  &$8.0\pm0.2$ 		    &$-375^{+28}_{-28}$\\ 
A 1651 		&$0.084$  &$8.4\pm0.7$ 		    &$-247^{+30}_{-30}$\\ 
A 401 		&$0.074$  &$6.4\pm0.2$ 		    &$-338^{+20}_{-20}$\\ 
A 399 		&$0.072$  &$9.1\pm0.4$ 		    &$-164^{+21}_{-21}$\\ 
A 2256 		&$0.058$  &$9.7\pm0.8$ 		    &$-243^{+29}_{-29}$\\ 
A 1656 		&$0.023$  &$6.6\pm0.2$ 		    &$-302^{+48}_{-48}$\\
\enddata
\tablecomments{Clusters in the sample; their redshift, gas temperature and central temperature decrement.}
\label{tab:sample}
\end{deluxetable}

\section{X-Ray Morphology in Two Dimensions}
\label{sec:morph_2D}
{\it Chandra} and {\it XMM}
observations of clusters in the past few years have shown that in 
general clusters exhibit elliptical surface brightness maps, and so 
cannot be spherically symmetric.  In order to obtain a uniform set of 
X-Ray observables for our sample objects, we have re-analyzed archival
X-Ray data for each of them.   We have used {\it Chandra} and/or {\it XMM}
data for all objects except  A~520, for which only {\it ROSAT} data 
are available.\\
We modeled the emission of all clusters in the $0.3-7.0\ {\rm keV}$
band. Pixel values of all detected point sources were replaced with 
values interpolated from the surrounding background regions; 
the {\it CIAO} tools {\it wavdetect} and {\it dmfilth} were used for this purpose.\\
Using the {\it SHERPA}
software, we fitted the cluster surface brightness to elliptical
2-D $\beta$-models (see Eq.~\ref{eq:sxb1}). Results are
listed in Table~\ref{tab:2D_fit}. Fitted models from {\it Chandra} and {\it XMM}
observations are roughly consistent.\\
The 25 clusters have a weighted median projected axis ratio of $e_{\rm proj}=1.24 {\pm} 0.09$, 
in very good agreement with the value of $\langle e_{\rm proj}\rangle =1.25 {\pm}0.19$ obtained by~\cite{Moh95} from {\it Einstein} data of a lower-redshift sample of 65 objects.
Only six of the $25$ clusters have a projection in the p.o.s. close to
be circular ($e_{\rm proj} <1.2$).

\tabletypesize{\scriptsize}
\def\arraystretch{1.0}
\begin{deluxetable*}{lrrrrrrc}
\tablecolumns{7}
\tablewidth{0pt}
\tablecaption{Two-Dimensional Analysis}
\tablehead{
\colhead{}  &\multicolumn{2}{c}{$x_{\rm c},y_{\rm c}$} &\colhead{}  &\colhead{}  &\colhead{} &\colhead{}    &\colhead{}\\
\colhead{Cluster}  &\colhead{R.A.}&\colhead{Decl.} &\colhead{$e_{\rm proj}$}  &\colhead{$\theta$}  &\colhead{$r_c$} &\colhead{$\beta$}    &\colhead{Satellite}\\
\colhead{} &\colhead{}         &\colhead{}&\colhead{}            &\colhead{(deg)}     &\colhead{(arcsec)}  &\colhead{}  &\colhead{}
}
\startdata
MS 1137.5+6625  &$11\ 40\ 22.3$ & $+66\ 08\ 15.3$  &$1.113\pm0.014$ & $63.9\pm1.0$&  $11.28\pm0.55$  & $0.58\pm0.01$ & $1$\\
MS 0451.6-0305  &$04\ 54\ 11.4$ & $-03\ 00\ 51.3$  &$1.307\pm0.015$ & $84.1\pm1.1$&  $45.1\pm1.2$  & $0.88\pm0.02$ & $1$\\
Cl 0016+1609    &$00\ 18\ 33.5$ & $+16\ 26\ 12.9$  &$1.205\pm0.013$ & $310.8\pm1.7$& $36.42\pm0.93$  & $0.63\pm0.01$ & $1$\\
Cl 0016+1609    &$00\ 18\ 33.1$ & $+16\ 26\ 10.5$  &$1.168\pm0.019$ & $314\pm3$  	&  $39.0\pm1.1$  & $0.65\pm0.08$ & $2$\\
RXJ1347.5-1145  &$13\ 47\ 30.7$ & $-11\ 45\ 09.1$  &$1.453\pm0.019$ & $21.5\pm1.0$&  $6.36\pm0.15$  & $0.533\pm0.003$ & $1$\\
A 370           &$02\ 39\ 53.3$ & $-01\ 34\ 39.0$  &$1.564\pm0.018$ & $353.8\pm0.7$&  $50.1\pm1.8$  & $0.52\pm0.01$ & $1$\\
MS 1358.4+6245  &$13\ 59\ 50.7$ & $+62\ 31\ 04.1$  &$1.325\pm0.019$ & $23.4\pm1.4$&  $14.23\pm0.49$  & $0.526\pm0.004$ & $1$\\
A 1995          &$14\ 52\ 57.9$ & $+58\ 02\ 55.8$  &$1.242\pm0.010$ & $122.2\pm1.0$&  $45.78\pm0.88$  & $0.73\pm0.01$ & $1$\\
A 611           &$08\ 00\ 56.8$ & $+36\ 03\ 23.5$  &$1.14\pm0.05$ & $326\pm9$  	&  $21.84\pm0.62$  & $0.596\pm0.006$ & $1$\\
A 697           &$08\ 42\ 57.6$ & $+36\ 21\ 56.8$ &$1.334\pm0.016$ & $16.2\pm1.2$ &  $54.3\pm1.7$  & $0.64\pm0.01$ & $1$\\
A 1835          &$14\ 01\ 02.0$ & $+02\ 52\ 42.9$  &$1.225\pm0.012$ & $7.0\pm1.4$ &  $8.34\pm0.14$  & $0.511\pm0.002$ & $1$\\
A 2261          &$17\ 22\ 27.1$ & $+32\ 07\ 57.4$  &$1.022\pm0.017$ & $0.0\pm1.7$ &  $20.58\pm0.75$  & $0.578\pm0.007$ & $1$\\
A 773           &$09\ 17\ 53.1$ & $+51\ 43\ 37.9$ &$1.237\pm0.022$ & $0.0\pm2.5$  &  $49.5\pm1.9$  & $0.63\pm0.02$ & $1$\\
A 773           &$09\ 17\ 52.7$ & $+51\ 43\ 37.0$  &$1.184\pm0.019$ & $0.0\pm2.9$ &  $45.2\pm2.3$  & $0.58\pm0.06$ & $2$\\
A 2163          &$16\ 15\ 46.6$ & $-06\ 08\ 44.9$  &$1.206\pm0.004$ & $0.0\pm0.6$ &  $94.80\pm0.95$  & $0.720\pm0.005$ & $1$\\
A 2163          &$16\ 15\ 46.0$ & $-06\ 08\ 46.9$  &$1.163\pm0.009$ & $0.0\pm1.4$ &  $91.3\pm1.3$  & $0.71\pm0.01$ & $2$\\
A 520           &$04\ 54\ 09.8$ & $+02\ 55\ 22.4$  &$1.06\pm0.05$ & $347\pm20$   	&  $115.2\pm11.8$  & $0.59\pm0.04$ & $3$\\
A 1689          &$13\ 11\ 29.6$ & $-01\ 20\ 28.0$  &$1.141\pm0.012$ & $342.4\pm2.2$&  $27.01\pm0.73$  & $0.582\pm0.005$ & $1$\\
A 665           &$08\ 30\ 57.1$ & $+65\ 51\ 01.8$  &$1.238\pm0.012$ & $33.7\pm1.2$&  $63.0\pm1.3$  & $0.582\pm0.006$ & $1$\\
A 2218          &$16\ 35\ 51.9$ & $+66\ 12\ 34.6$  &$1.162\pm0.009$ & $83.5\pm1.5$&  $56.0\pm0.84$  & $0.591\pm0.004$ & $1$\\
A 2218          &$16\ 35\ 52.4$ & $+66\ 12\ 33.5$  &$1.201\pm0.013$ & $85.4\pm1.7$&  $58.5\pm1.1$  & $0.603\pm0.009$ & $2$\\
A 1413          &$11\ 55\ 17.9$ & $+23\ 24\ 16.2$  &$1.473\pm0.019$ & $182.2\pm0.9$&  $43.0\pm1.4$  & $0.573\pm0.007$ & $1$\\
A 2142          &$15\ 58\ 20.1$ & $+27\ 14\ 03.5$  &$1.540\pm0.007$ & $52.3\pm0.3$&  $73.1\pm1.1$  & $0.598\pm0.005$ & $1$\\
A 478           &$04\ 13\ 25.3$ & $+10\ 27\ 53.5$  &$1.477\pm0.006$ & $43.5\pm0.3$&  $30.53\pm0.16$  & $0.533\pm0.001$ & $1$\\
A 1651          &$12\ 59\ 21.9$ & $-04\ 11\ 44.6$  &$1.184\pm0.013$ & $272.4\pm1.9$&  $73.4\pm2.0$  & $0.598\pm0.007$ & $1$\\
A 401           &$02\ 58\ 57.1$ & $+13\ 34\ 37.8$  &$1.303\pm0.008$ & $325.1\pm0.6$&  $142.8\pm3.0$  & $0.625\pm0.007$ & $1$\\
A 399           &$02\ 57\ 52.0$ & $+13\ 02\ 38.7$  &$1.207\pm0.009$ & $337.5\pm1.2$&  $139.8\pm3.8$  & $0.536\pm0.008$ & $2$\\
A 2256          &$17\ 04\ 00.4$ & $+78\ 38\ 37.1$  &$1.327\pm0.008$ & $61.3\pm0.5$&  $529\pm16$  & $1.33\pm0.06$ & $2$\\
A 1656    	&$12\ 59\ 44.1$ & $+27\ 56\ 43.0$  &$1.141\pm0.006$ & $0.0\pm0.6$ &  $540\pm34$  & $0.65\pm0.07$ & $2$\\
\enddata
\tablecomments{Fit parameters of the elliptical $\beta$ model: $x_{\rm c},y_{\rm c}$ is the central position; $e_{proj}$ is the projected axial ratio; $\theta$ is the orientation angle (north over east); $r_c$ and $\beta$ are the model core radius and slope, respectively. In the last column, label $1$ is for {\it Chandra}, $2$ for {\it XMM} and $3$ for {\it ROSAT} HRI observation.}
\label{tab:2D_fit}
\end{deluxetable*}

\subsection{Circular Versus Elliptical $\beta$ Models}
\label{sec:sph_vs_ell}
\begin{figure}
\epsscale{0.6}
\plotone{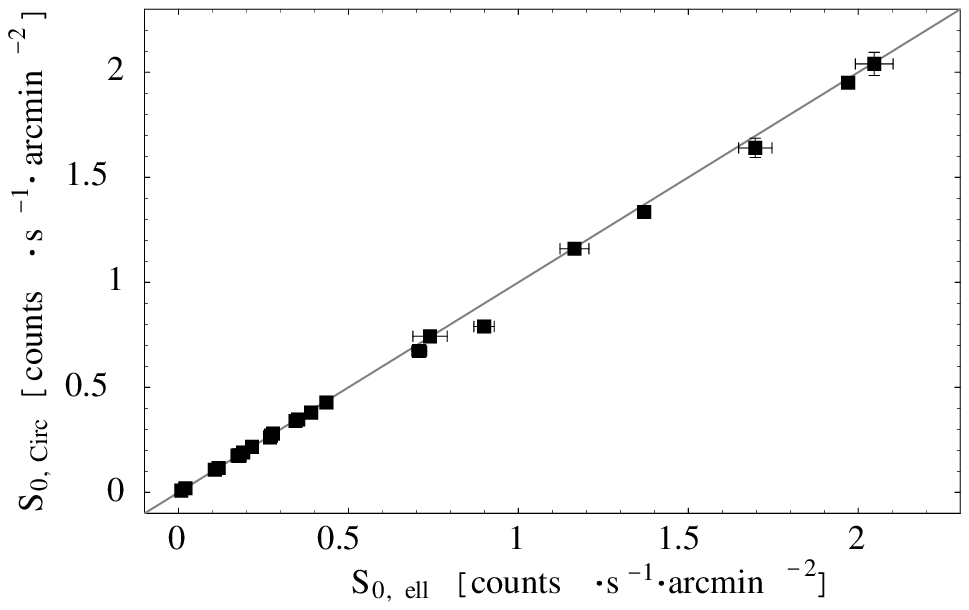}
\plotone{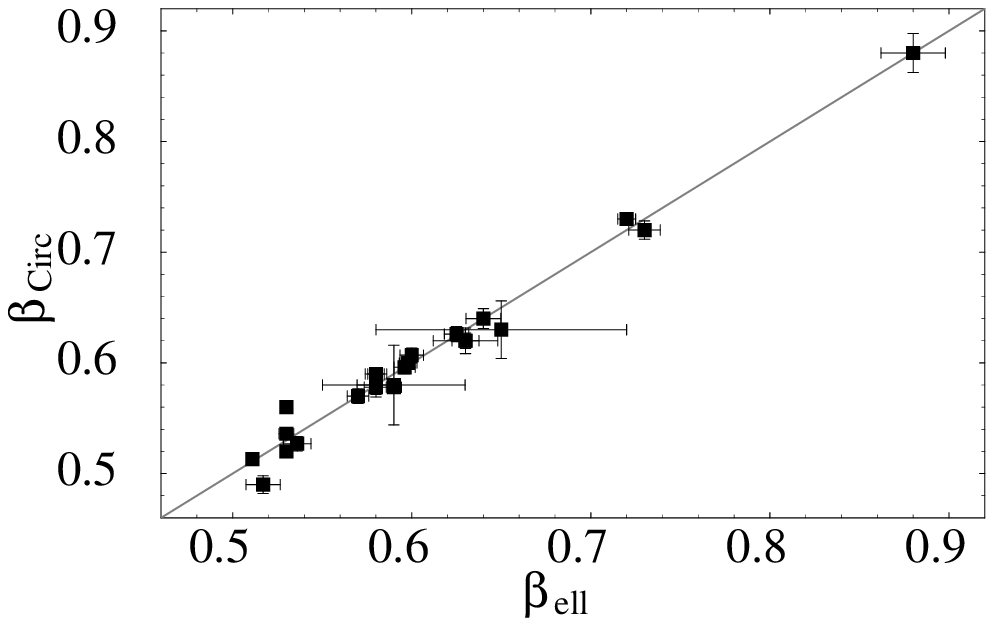}
\plotone{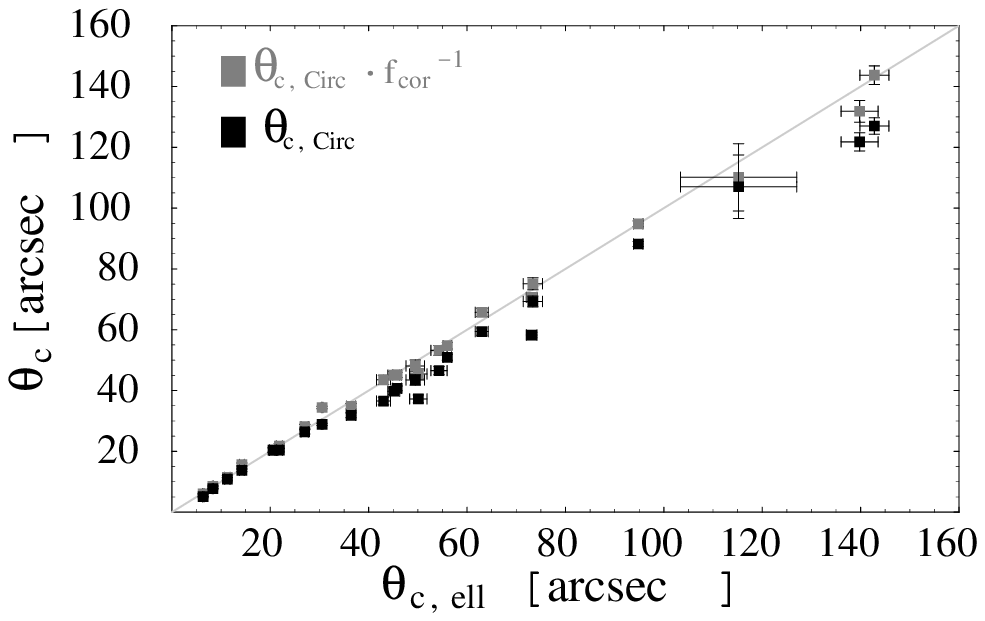}
\caption{Comparison of central surface brightness {\it (top)}, slope $\beta$ {\it (middle)} 
and  core radius {\it (bottom)} obtained fitting circular and elliptical 
$\beta$-models to the X-Ray surface brightness maps of sample clusters.  
In the  bottom panel the gray symbols show core radii corrected for the ellipticity as in Eq.~\ref{eq:thetac_corr}.}
\label{fig:correzioni}
\end{figure}

Although clusters are rarely circular in projection, some previous 
joint analyses of X-Ray and SZE data have assumed spherical symmetry. 
In order to bound the effects of this simplification, 
we have also modeled the surface brightness profiles of 
each  sample cluster  with a circular $\beta$-model.
The choice of circular rather than elliptical $\beta$ model does not affect 
the resulting  of the central surface brightness, as shown in the top 
panel in Fig.~\ref{fig:correzioni}. 
For a few clusters the fitted value of the slope $\beta$ differs slightly
between circular and  elliptical models (middle panel of 
Fig.~\ref{fig:correzioni}). 
As would be expected, however, significantly different values for the
core radius are obtained with these two models (bottom panel of Fig. ~\ref{fig:correzioni}).  This behavior has already been noted  by~\cite{Hug98}.\\
 Therefore, relaxing the assumption of  circular projection on the p.o.s. 
when measuring the angular diameter distance (Eq.~\ref{eq:obl7}), mainly affects the value of the 
projected core radius $\theta_{c,\rm proj}$. The bottom panel in Fig.~\ref{fig:correzioni} 
shows that the core radius obtained using a circular $\beta$-model ($\theta_{c,\rm Circ}$)
is consistently lower (black squares) than that obtained from an elliptical 
model ($\theta_{c,\rm Ell}$). $\theta_{c,\rm Circ}$ can in fact be well approximated
by the arithmetic mean of the two semi-axes of the elliptical isophotes in the p.o.s. \\
The angular diameter distance obtained assuming spherical symmetry
(Table~\ref{tab:Dc}) can therefore, in first approximation, be corrected 
for the 
observed ellipticity of the cluster in the p.o.s. multiplying $\left. D_{\rm c} \right|_{\rm Exp}^{\rm Circ}$ 
by the correction factor $f_{\rm cor}$:
\begin{equation}
\left. D_{\rm c} \right|_{\rm Exp}^{\rm Circ,Corr} =\left. D_{\rm c}\right|_{\rm Exp}^{\rm Circ} f_{\rm cor} = \left. D_{\rm c}\right|_{\rm Exp}^{\rm Circ}\frac{1+e_{\rm proj}}{2 e_{\rm proj}}.
\label{eq:corr_factor}
\end{equation}
As shown in the bottom panel of Fig.~\ref{fig:correzioni}, the corrected values of the 
core radii  $\left. \theta_{\rm c}\right|_{\rm Circ}^{\rm Cor}$ 
\begin{equation}
\left. \theta_{\rm c} \right|_{\rm Circ}^{\rm Cor}=\theta_{c,\rm Circ}\ \frac{1}{f_{\rm cor}}
\label{eq:thetac_corr}
\end{equation}
provide  a good approximation to $\theta_{c,\rm ell}$ (gray squares).

\section{Cluster Morphology in Three Dimensions}
\label{sec:tria}
\subsection{Angular Diameter Distances}
\label{sec:Dc_Ell}
\begin{figure}
\epsscale{1.0}
\plotone{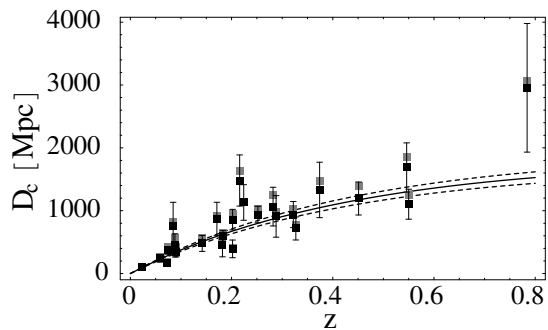}
\caption{$\left. D_{\rm c}\right|_{\rm Cosm}$ as a function of redshift (solid line); dashed lines represent the 1$\sigma$ confidence level. Black and grey point represent $\left. D_{\rm c}\right|_{\rm Exp}^{\rm Ell}$ and $\left. D_{\rm c}\right|_{\rm Exp}^{\rm Circ}$ computed in this work, respectively.} 
\label{fig:Dc_vs_z}
\end{figure}

In order to estimate the l.o.s. extent of clusters, then, we need 
only to obtain values of $\left. D_{\rm c}\right|_{\rm Exp}$ from the X-Ray and 
SZE data (via Eq.~\ref{eq:obl7}) and compare them (via Eq.~\ref{eq:simp3}) 
to the angular size distance obtained from the measured redshift  
and our adopted cosmological model.  Since only the X-Ray data are publically 
available, however, we are unable to jointly fit both SZE 
and X-Ray data. For this reason we must rely on published values of 
central CMB temperature decrement ($\Delta T_{0}$) for our analysis.\\
A potential difficulty with this approach is that the 
available values of $\Delta T_{0}$,  from~\cite{Ree02} and~\cite{Mas01}, 
have been inferred assuming that  clusters are circularly symmetric 
when seen in projection on the sky.  While this assumption is quite 
reasonable given the limited spatial resolution of the data available to 
these authors, it is not, in general, consistent with the results of our 
analysis of higher-resolution X-Ray data (see Table~\ref{tab:2D_fit}). 
In the limit of very high spatial resolution SZE data, we would expect this
inconsistency to have negligible effect on our results, just as we find that
with high-resolution X-Ray data, the same  central X-Ray surface 
brightness is inferred from fits of circular and elliptical models 
(see the top panel in Fig.~\ref{fig:correzioni}). \\
We have computed values of the angular diameter distances for all clusters in the sample 
both under the assumption of spherical symmetry, and relaxing the assumption to a more 
general triaxial morphology. For the spherical case results obtained modeling the cluster X-Ray surface
brightness profiles with circular $\beta$-models (\S~\ref{sec:sph_vs_ell}) were substituted  
into Eq.~(\ref{eq:obl7}). For the triaxial case results from the elliptical $\beta$-models 
were instead used (\S~\ref{sec:morph_2D}).
The term $f(\nu,T_e)$, which accounts for the frequency shift and also includes relativistic corrections,
was computed as described by~\cite{Ito98}.
Central values of the CMB temperature decrement ($\Delta T_0$) were taken from~\cite{Ree02} 
and~\cite{Mas01}.
The resulting values of $\left. D_{\rm c} \right|_{\rm Exp}^{\rm Ell}$ are listed in Table~\ref{tab:Dc}.
The largest source of error (about $70\%$ of the total) is the uncertainty 
on the SZE measurement of $\Delta T_0$. The second most significant error 
source is the uncertainty in the X-Ray measurement of the intra-cluster 
plasma temperature (about $20\%$). 
Both $\left. D_{\rm c} \right|_{\rm Exp}^{\rm Ell}$ and $\left. D_{\rm c} \right|_{\rm Exp}^{\rm Circ}$
are plotted in Fig.~\ref{fig:Dc_vs_z}. A comparison between the experimental quantities $\left. D_{\rm c} \right|_{\rm Exp}^{\rm Circ}$ and $\left. D_{\rm c} \right|_{\rm Exp}^{\rm Ell}$ and the values of $\left. D_{\rm c} \right|_{\rm Cosm}$, together with their relative errors, highlights the high precision to which the cosmological angular distance is known, compared to the two experimental estimates, in support of our approach of a fixed cosmological model.

\tabletypesize{\scriptsize}
\def\arraystretch{1.0}
\begin{deluxetable}{lrrr}
\tablecolumns{7}
\tablewidth{0pt}
\tablecaption{Angular Diameter Distance}
\tablehead{
\colhead{Cluster} &\colhead{$\left. D_{\rm c}\right|_{\rm Cosm}$} &\colhead{$\left. D_{\rm c}\right|_{\rm Exp}^{\rm Circ}$}  &\colhead{$\left. D_{\rm c}\right|_{\rm Exp}^{\rm Ell}$}\\
\colhead{}   &\colhead{(Mpc)}   &\colhead{(Mpc)} &\colhead{(Mpc)}
}
\startdata
MS 1137.5+6625  &$1537^{+71}_{-91}$&$3179^{+1103}_{-1640}$&$2479\pm1023$\\ 
MS 0451.6-0305  &$1322^{+58}_{-76}$&$1278^{+265}_{-299}$  &$1073\pm238$  \\ 
Cl 0016+1609    &$1318^{+58}_{-76}$&$2041^{+484}_{-514}$  &$1635\pm391$  \\ 
RXJ1347.5-1145	&$1189^{+52}_{-68}$&$1221^{+368}_{-343}$  &$1166\pm262$     \\ 
A 370 		&$1063^{+46}_{-61}$&$4352^{+1388}_{-1245}$&$1231\pm441$\\ 
MS 1358.4+6245	&$974^{+41}_{-55}$&$866^{+248}_{-310}$   &$697\pm183$   \\ 
A 1995		&$964^{+41}_{-54}$&$1119^{+247}_{-282}$  &$885\pm207$  \\ 
A 611		&$893^{+38}_{-50}$&$995^{+325}_{-293}$   &$934\pm331$   \\ 
A 697 		&$880^{+37}_{-49}$&$998^{+298}_{-250}$   &$1099\pm308$   \\ 
A 1835 		&$811^{+37}_{-45}$&$1027^{+194}_{-198}$  &$946\pm131$  \\
A 2261 		&$743^{+31}_{-41}$&$1049^{+306}_{-272}$  &$1118\pm283$  \\ 
A 773 		&$722^{+30}_{-40}$&$1450^{+361}_{-332}$  &$1465\pm407$  \\ 
A 2163 		&$686^{+29}_{-38}$&$828^{+181}_{-205}$   &$806\pm163$   \\ 
A 520 		&$686^{+29}_{-38}$&$723^{+270}_{-236}$   &$387\pm141$   \\ 
A 1689 		&$634^{+27}_{-35}$&$688^{+172}_{-163}$   &$604\pm84$   \\ 
A 665 		&$632^{+26}_{-35}$&$466^{+217}_{-179}$   &$451\pm189$   \\ 
A 2218 		&$601^{+25}_{-33}$&$1029^{+339}_{-352}$  &$809\pm263$  \\ 
A 1413 		&$515^{+21}_{-29}$&$573^{+171}_{-151}$   &$478\pm126$   \\ 
A 2142 		&$349^{+14}_{-19}$&$187^{+212}_{-97}$    &$335\pm70$    \\ 
A 478 		&$340^{+14}_{-19}$&$406^{+237}_{-135}$   &$448\pm185$   \\ 
A 1651 		&$327^{+14}_{-18}$&$373^{+202}_{-122}$   &$749\pm385$   \\ 
A 401 		&$289^{+12}_{-16}$&$610^{+593}_{-254}$   &$369\pm62$   \\ 
A 399 		&$282^{+12}_{-16}$&$107^{+85}_{-41}$     &$165\pm45$     \\ 
A 2256 		&$232^{+10}_{-13}$&$296^{+127}_{-90}$    &$242\pm61$    \\ 
A 1656 		&$96^{+4}_{-5}$&$235^{+218}_{-98}$    &$103\pm42$    \\
\enddata
\tablecomments{Cosmological angular diameter distance, its experimental estimate as reported by~\cite{Ree02} and
~\cite{Mas01} assuming spherical symmetry, and the experimental quantity $\left. D_{\rm c}\right|_{\rm Exp}^{\rm Ell}$, defined by Eq.~(\ref{eq:obl7}), computed in this paper assuming the clusters are oblate spheroids. }
\label{tab:Dc}
\end{deluxetable}

\subsection{Elongation Along the Line of Sight}
\label{sec:ellos}
\begin{figure}
\epsscale{1.0}
\plotone{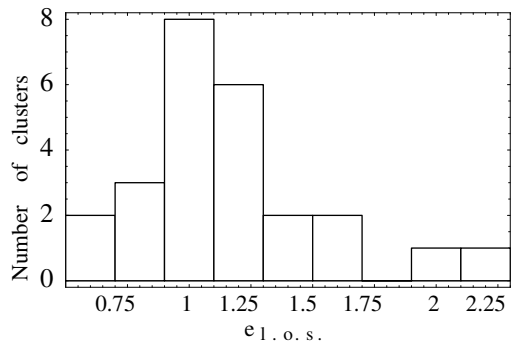}
\caption{Distribution of elongation along the l.o.s. for all clusters in our sample.}
\label{fig:isto_e_los}
\end{figure}

Assuming a general triaxial morphology, the ratio between $\left. D_{\rm c}\right|_{\rm Exp}^{\rm Ell}$ 
and $\left. D_{\rm c}\right|_{\rm Cosm}$ provides an estimate of the ratio of the 
cluster axis along the l.o.s. and the cluster major axis in the p.o.s.
We have computed values of $\left. D_{\rm c}\right|_{\rm Exp}^{\rm Ell}$ for all clusters 
in the sample (\S~\ref{sec:Dc_Ell}). 
For each cluster we have then also computed $\left. D_{\rm c}\right|_{\rm Cosm}$ (Eq.~\ref{eq:crit3})
and have then estimated their $e_{\rm l.o.s.}$. Resulting values are listed in 
Table~\ref{tab:3D_morph}.\\
Since the observables in our analysis have asymmetric uncertainties, we apply corrections given by~\cite{Dag04} to obtain estimates of sample 
mean and standard deviation. \\
All clusters in our sample were X-Ray selected; X-Ray surveys are surface 
brightness limited. Clusters close to the detection limit which are elongated 
along the l.o.s. will be detected, while the ones which are more extended in 
the p.o.s. will be missed. If a surface brightness limit is fixed 
which is far above the detection limit of the survey, the problem should be eliminated.\\
In both the~\cite{Ree02} and the~\cite{Mas01} samples this ``correction'' limit 
was applied. Our final sample shows in fact only mild signs of preferential 
elongation of the clusters along the l.o.s. (see Fig.~\ref{fig:isto_e_los}).
 Of the $25$ clusters, $15$ clusters are in fact more elongated along the l.o.s 
($e_{\rm l.o.s.} > 1$), while the remaining $10$ clusters are compressed. 
The mean of the distribution of the elongations is 
$\langle e_{\rm l.o.s.} \rangle = 1.15 {\pm} 0.08$. 
In presence of likely outliers, the median is a more stable estimator~\citep{Got01}. 
The median of the $e_{\rm l.o.s.}$'s is $1.08\pm 0.17$. 
While on average we observe only a very slight preferential elongation of the clusters along 
the l.o.s., residual of X-Ray selection effects, clusters with extreme axes
ratios are still preferentially selected if the elongation lies along 
the l.o.s. This is a clear example of how deeply X-Ray selected cluster 
samples are affected by morphological and orientation issues.

\tabletypesize{\scriptsize}
\def\arraystretch{1.0}
\begin{deluxetable}{lrr}
\tablecolumns{7}
\tablewidth{0pt}
\tablecaption{3-D Morphology}
\tablehead{
\colhead{Cluster}&\colhead{$e_{\rm l.o.s.}$}&\colhead{q$_{\rm max}$}  \\
}
\startdata
MS 1137.5+6625   &$1.61\pm{0.69}$  &$1.79\pm{0.68}$ \\ 
MS 0451.6-0305   &$0.81\pm{0.18}$& $1.31\pm{0.11}$ \\ 
Cl 0016+1609	 &$1.24\pm{0.30}$& $1.49\pm{0.32}$ \\
RXJ1347.5-1145   &$0.98\pm{0.22}$  &$1.45\pm{0.19}$ \\
A 370            & $1.16\pm{0.42}$ & $1.81\pm{0.48}$ \\ 
MS 1358.4+6245	 & $0.72\pm{0.19}$ & $1.40\pm{0.12}$ \\ 
A 1995		 & $0.92\pm{0.21}$ & $1.24\pm{0.14}$ \\ 
A 611		 & $1.05\pm{0.37}$ & $1.19\pm{0.27}$ \\ 
A 697		 & $1.25\pm{0.35}$ & $1.67\pm{0.40}$ \\ 
A 1835		 & $1.17\pm{0.16}$ & $1.43\pm{0.18}$ \\ 
A 2261		 & $1.51\pm{0.38}$ & $1.54\pm{0.39}$ \\ 
A 773		 & $2.03\pm{0.56}$ & $2.51\pm{0.71}$ \\ 
A 2163		 & $1.18\pm{0.24}$ & $1.37\pm{0.23}$ \\ 
A 520		 & $0.56\pm{0.20}$ & $1.77\pm{0.26}$ \\ 
A 1689		 & $0.95\pm{0.13}$ & $1.14\pm{0.08}$ \\ 
A 665		 & $0.71\pm{0.30}$ & $1.40\pm{0.29}$ \\ 
A 2218		 & $1.85\pm{0.44}$ & $1.57\pm{0.45}$ \\ 
A 1413		 & $0.93\pm{0.25}$ & $1.47\pm{0.20}$ \\ 
A 2142		 & $0.96\pm{0.20}$ & $1.54\pm{0.18}$ \\ 
A 478		 & $1.32\pm{0.54}$ & $1.95\pm{0.65}$ \\ 
A 1651		 & $2.29\pm{1.18}$ & $2.71\pm{1.39}$\\ 
A 401		 & $1.28\pm{0.21}$ & $1.67\pm{0.27}$ \\ 
A 399		 & $0.58\pm{0.16}$ & $1.71\pm{0.12}$ \\ 
A 2256		 & $1.04\pm{0.26}$ & $1.38\pm{0.22}$ \\ 
A 1656		 & $1.08\pm{0.43}$ & $1.23\pm{0.32}$ \\
\enddata
\tablecomments{Cluster elongation along the l.o.s. and maximum axial ratio.}
\label{tab:3D_morph}
\end{deluxetable}

\subsection{Maximum Axis Ratio}
\label{sec:qmax}
\begin{figure}
\epsscale{1.0}
\plotone{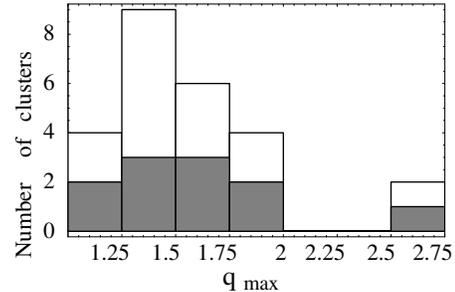}
\caption{Distribution of maximum axial ratios.
The gray histogram is for cooling-flow systems.}
\label{fig:isto_q_max}
\end{figure}

\begin{figure}
\epsscale{1.0}
\plotone{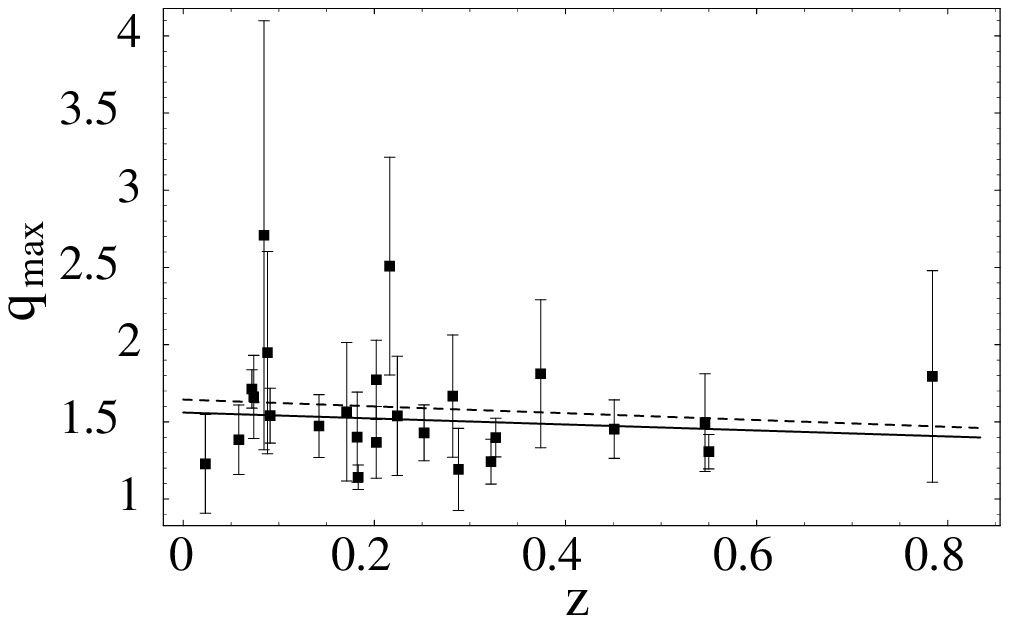}
\caption{Maximum axial ratio as a function of the redshift. The solid and dashed lines represent the linear best fit to the data with and without weights, respectively.}
\label{fig:qMax_vs_z}
\end{figure}

\begin{figure}
\epsscale{1.0}
\plotone{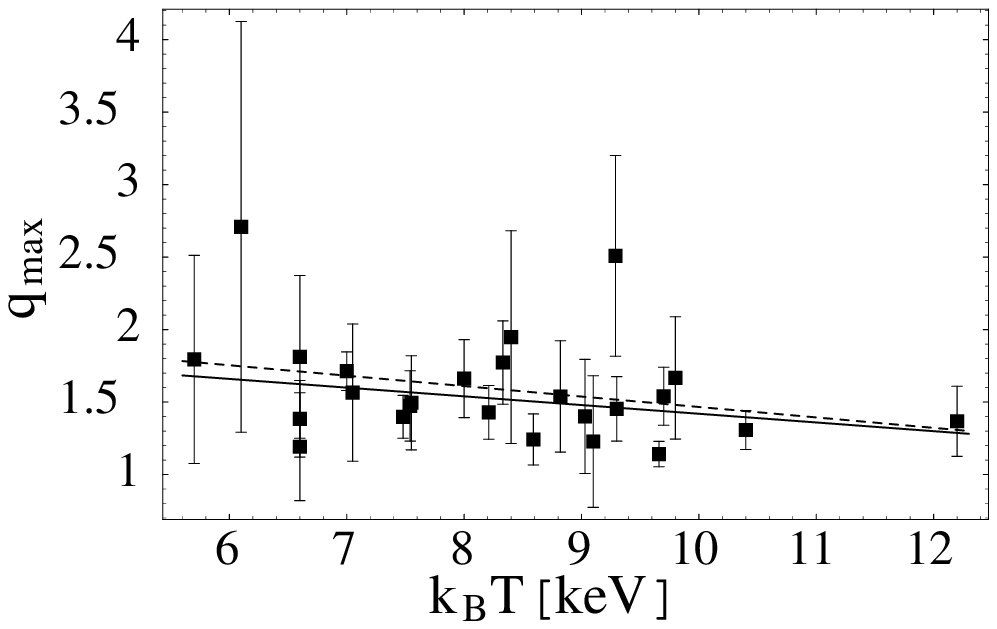}
\caption{Maximum axial ratio as a function of temperature. The solid and dashed lines 
represent the linear best fit to the data with and without weights,
respectively.}
\label{fig:qMax_vs_T}
\end{figure}

\begin{figure}
\epsscale{1.0}
\plotone{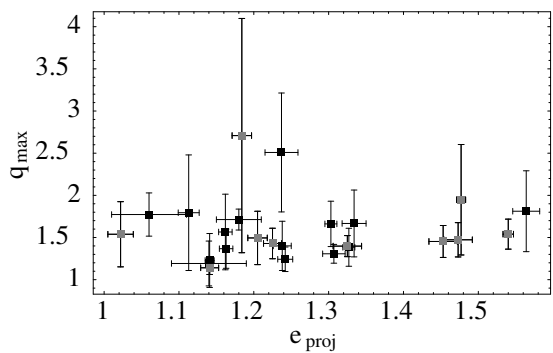}
\caption{Cluster maximum axial ratio as 
function of (projected) ellipticity in the p.o.s. 
Grey squares denote cooling-flow clusters.}
\label{fig:qMax_vs_eHB}
\end{figure}

We can estimate the three ellipsoidal axis lengths ($v_1$, $v_2$ and $v_3$ )
from the measured values of $e_{\rm l.o.s.}$ and
$e_{\rm proj}$, and from these the  ratio of  the semi-major to the 
semi-minor axis, $q_{\rm max}$. 
$q_{\rm max}$ is an extremely convenient tool to describe the intrinsic shape of a cluster since it allows, without the aid of further parameters, to quantify how far a cluster is from spherical symmetry.
For most clusters in the sample, the confidence regions of $v_1$, $v_2$ 
and $v_3$ are highly overlapping so that, for example, the upper bound of
the 1-$\sigma$ interval for the estimate of the ratio between the median
and the minor axis, $q_{\rm mid}$, may be larger than the upper limit of 
$q_{\rm max}$.\\
To obtain well defined estimates of the errors of the maximum,
intermediate and minimum
axis ratios for each cluster, assuming the $v_i$'s to be normally
distributed, we have obtained  $10^5$ random samples from each distribution.
We have then selected the
maximum, the intermediate and the minimum values of each set of three in
order to build the distribution of the maximum, intermediate and minimum
axis ratios. We have finally computed the standard
deviations of such three distributions, that provide estimates for the
errors for the axes ratios.
The resulting values of $q_{max}$ are listed in Table~\ref{tab:3D_morph} and
their distribution is shown in Fig.~\ref{fig:isto_q_max}. \\
$q_{\rm max}$ has a mean value of $\langle
q_{\rm max}\rangle = 1.59 {\pm} 0.07$, and a median of $1.49\pm0.17$.
This result is consistent with cosmological simulations in which the
mean value of the maximum axial ratio ranges from $1.56$~\citep{Suw03,Kas04} 
to $1.8$~\citep{Jin02}. The intermediate axis ratios, $q_{\rm mid}$ show a median of $1.21\pm0.12$.
At $1-\sigma$ no cluster in the sample can be approximated as
spherical;  $11$ clusters are spherical at the $3-\sigma$ confidence level.\\
Although our estimates are affected by  large errors and the data
sample is of modest size, we look for trends in the distribution of
the maximum axial ratios. \\
No correlation is observed between the maximum axial ratio and redshift
(see Fig.~\ref{fig:qMax_vs_z} where the solid and dashed lines represent the 
weighted and non weighted linear best fit to the data, respectively).\\
A poor correlation is observed also between the maximum axial ratio and 
the cluster gas temperature.
We find at most a weak tendency for hotter clusters to exhibit 
smaller values of $q_{max}$. The linear weighted best fit to the data is: 
$q_{\rm max}=(2.02{\pm}0.35)-(0.060 {\pm}0.041)\ T$.
The trend is plotted in Fig.~\ref{fig:qMax_vs_T}, where the solid and 
dashed lines represent the weighted and non weighted linear best fit 
to the data, respectively.  The absence of such a correlation may 
indicate that, in our sample, high cluster temperatures are not 
predominantly the result of shocks associated with accretion of 
sub-clusters~\cite{Ran02}, since such accretion
events seem likely also to produce departures from spherical morphology.\\
From the distribution of axial shapes of clusters in our sample, 
we can estimate the effect that the assumption of spherical symmetry 
has on the determination of the total cluster mass.
If the mass is computed at large distances from the cluster center 
($\geq 1\ {\rm Mpc}$) the difference between the two models is less than 
$2 \%$ even for the most elongated clusters. If the mass is computed 
close to the cluster core the effect becomes larger,  
ranging from $4 \%$ to $25 \%$, for less to more elongated clusters in 
our sample, respectively, when the mass is computed 
within a sphere of radius $100\ {\rm kpc}$.
Triaxial cluster distributions could therefore
at least partially account for the observed discrepancies in the total mass
of clusters computed with lensing and X-Rays measurements.\\
We then analyze a subsample of the $10$ clusters for which the
presence of a cooling flow has been claimed (i.e. for which the upper limit, $90 \%$ confidence, to the central cooling time has been measured to be less than $10^{10}\ {\rm yr}$). Cooling flow clusters are
typically recognized as dynamically relaxed systems in which the ICM is
supported by thermal pressure which dominates over non-thermal
processes.  Their X-Ray emission is in most cases regular and symmetric
and little or no substructures is visible at optical wavelengths. 
We find no indication, however, that cooling flow clusters are more likely
to be spherical.    Fig.~\ref{fig:isto_q_max} suggests 
that the distribution of  maximum axial ratios for the cooling flow sample is 
indistinguishable from that of the sample as a whole; 
a Kolmogorov-Smirnov test confirms this impression.\\
Finally, we find no relationship between 
cluster elongation along the l.o.s. and 2-D ellipticity 
(see Fig.~\ref{fig:qMax_vs_eHB}).  In particular, a circular 
(projected) surface brightness profile is not an indicator 
that a cluster is in fact spherical.

\subsection{Ellipticity and Prolateness}
\begin{figure}
\epsscale{1.0}
\plotone{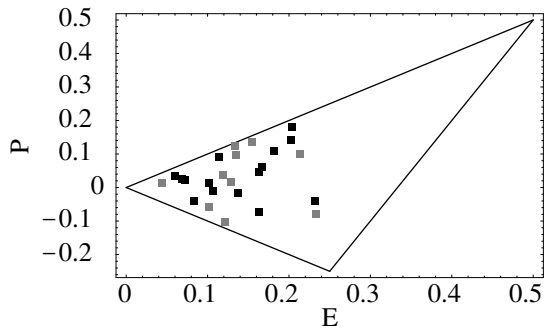}
\caption{Distribution in the ellipticity-prolateness plane for our sample.
Grey squares highlight cooling-flow clusters.}
\label{ell-pro}
\end{figure}

Triaxial ellipsoids can be represented in the ellipticity-prolateness plane
$(E,P)$~\citep{Tho98,Sul99}. The ellipticity is defined as:
\begin{equation}
\label{tria1}
E = \frac{1}{2}\frac{e_{\rm max}^2-e_{\rm min}^2}{e_{\rm min}^2 +
e_{\rm med}^2 + e_{\rm max}^2}
\end{equation}
and the prolateness as:
\begin{equation}
\label{tria2}
P = \frac{1}{2}\frac{e_{\rm min}^2 -2 e_{\rm med}^2 + e_{\rm
max}^2}{e_{\rm min}^2 + e_{\rm med}^2 + e_{\rm max}^2}
\end{equation}
where the axial ratios satisfy $e_{\rm min} \leq e_{\rm med} \leq e_{\rm
max}$. The allowed region in the $(E,P)$ plane is a triangle delimited
by the lines on which prolate and oblate clusters fall ($P=-E$ and
$P=E$, respectively) and the line connecting their endpoints.
Fig.~\ref{ell-pro} shows the distribution in ellipticity-prolateness
for our sample. No cluster in our sample shows extreme values
of the ellipticity parameter.  As expected from  
some simulations~\citep{Kas04}, prolate shapes ($P>0$) may be  more 
likely ($18$ clusters) than oblate ones  ($P<0$).
Once again, cooling flow clusters (gray boxes) are indistinguishable from the 
sample as a whole.

\section{Summary and Discussion}
\label{sec:disc}
In this paper we have discussed how observations of clusters in the microwave 
and X-Ray spectral bands can be combined to constrain their intrinsic 
3-D shapes, provided that the cosmological model is known. 
We have applied our method  to a combined sample of $25$ clusters of galaxies.
In doing so we make the simplifying assumption that the clusters are ellipsoids
with one axis parallel to the l.o.s. \\
Our sample clusters were originally selected on the basis of X-Ray luminosity, 
with the selection threshold well above the detection limit, in order to 
avoid selection on the basis of X-Ray surface brightness.
Even so, we observe that clusters with  extreme axes ratios are still 
preferentially selected only if the elongation lies along the l.o.s. \\
The mean value of the axial ratio in low-density cosmological simulations
ranges from $1.56$~\citep{Suw03,Kas04} to $1.8$~\citep{Jin02} and
$2.0$~\citep{Tho98}. This is consistent with results we present here
$\langle q_{\rm max} \rangle =1.59 {\pm} 0.07$. The spherical hypothesis is 
generally rejected, with prolate-like shapes being slightly more likely 
than oblate-like ones. 
Numerical investigations suggest there should be some tendency 
for axial ratios to be larger at higher redshift~\citep{Jin02,Suw03,Kas04}, 
though this effect is only marginal: the axial ratio increases 
only  $3\%$ from $z=0$ to
$z=0.8$~\citep{Kas04}. Our data are not sufficiently precise to test this
prediction. A poor correlation is also observed between the maximum axial ratio and 
the cluster gas temperature. The absence of such a correlation may 
indicate that high cluster temperatures are not mainly the result of shocks associated with accretion of 
sub-clusters, since such events would likely produce departures from spherical morphology.\\
The uncertainties on our results are mainly due to the relatively 
large errors  in the SZE measurements of central temperature decrement,
together with the quadratic dependence of the distance estimates on 
this parameter. More accurate SZE measurements, with better effective angular 
resolution,  are required to  extend this analysis to a large
sample spanning a greater redshift range.\\
A relevant number of cosmological tests are today based on the knowledge of the mass of galaxy clusters through X-Ray measurements; these masses are though usually computed assuming a spherical symmetry.
It is therefore extremely important to assess the effect that such assumption 
has on the determination of the total cluster mass.
We have estimated that while the effect is negligible when the mass is computed at large distances from the cluster center, the discrepancy between the two mass values becomes much more important as we get closer to the cluster core (i.e. $\approx 25 \%$  when the mass is computed 
within a radius of $100\ {\rm kpc}$).  
Triaxial cluster shapes may therefore
at least partially account for the discrepancies between cluster  mass
computed with strong lensing and X-Rays data.\\
Although the presence of a cooling flow is often interpreted as a
sign of dynamical relaxation, the properties of the subsample of $10$
cooling flow clusters do not differ from those of the whole sample:
cooling flow clusters therefore do not show preferentially spherical 
morphologies.

\acknowledgements
This work has been supported by NASA grants NAS8-39073 and NAS8-00128.
This paper is based on observations obtained from the {\it Chandra} Data Archive, 
the {\it XMM}-Newton Science Archive and the {\it ROSAT} Public Data Archive; we gratefully
thank the {\it Chandra} X-Ray Observatory Science Center (operated for NASA by the Smithsonian 
Astrophysical Observatory), {\it XMM}-Newton Space Operation Centre (operated by ESA) and 
MPE for maintaining the archives active and running.

\appendix
\section{A. Triaxial Ellipsoids}
\label{sec:triaxial}
We consider a cluster electron density distribution described by an
ellipsoidal triaxial $\beta$-model. In a $\beta$-model, the electron density 
of the intra-cluster gas is
assumed to be constant on a family of similar, concentric, coaxial
ellipsoids. High resolution $N$-body simulations have shown the
asphericity of density profiles of dark matter halos and how such
profiles can be accurately described by concentric triaxial ellipsoids
with aligned axis: ellipsoidal $\beta$-models therefore provide a more
detailed description of relatively relaxed simulated halos, respect to
the conventional spherically symmetric model~\citep{Jin02}.\\
In a coordinate system relative to the cluster, we then describe the
cluster electron density as:
\begin{equation}
\label{eq:tri0}
n_e  =  n_{e0} \left( 1+ \frac{\sum_{i=1}^3 v_i^2 x_{i,\rm
int}^2}{r_{\rm c}^2} \right)^{-3\beta/2}
\end{equation}
where $\left\{ x_{i,\rm int} \right\}$, with $i=1,2,3$, define an
intrinsic orthogonal coordinate system centered on the cluster's
barycenter and whose coordinates are aligned with its principal axes;
$r_{\rm c}$ is the characteristic length scale of the distribution,
which in our case is defined as the core radius; along each axis,
$v_i$ is the inverse of the corresponding core radius in units of
$r_{\rm c}$; $n_{e0}$ is the central electron density. The electron
density distribution in Eq.~(\ref{eq:tri0}) is described by 5
parameters: $n_{e0}$, $\beta$, the axial ratios $e_1\equiv v_1/v_3$ and
$e_2\equiv v_2/v_3$, and the core radius $r_{\rm c3}=r_{\rm c} / v_3$
along $x_{3,\rm int}$:
\begin{equation}
n_e  = n_{e0} \left( 1+ \frac{ e_1^2 x_{1,\rm int}^2 +e_2^2 x_{2,\rm
int}^2+ x_{3,\rm int}^2}{r_{\rm c3}^2} \right)^{-3\beta /2}
\label{eq:tri1}
\end{equation}
To write the electron density distribution given by
Eq.~(\ref{eq:tri1}) in a coordinate system relative to the observer,
three additional parameters are needed: the rotation angles
-$\theta_{\rm Eu}, \varphi_{\rm Eu}$ and $\psi_{\rm Eu}$- of the three
principal cluster axes respect to the observer. A rotation through the
first two Euler angles is sufficient to align the $x_{3,\rm obs}$-axis
of the observer coordinates system $\left\{x_{i,\rm obs}
\right\}$ with the l.o.s. of the observer, i.e. the
direction connecting the observer to the cluster center. When viewed
from an arbitrary direction, quantities constant on similar ellipsoids
project themselves on similar ellipses~\citep{Sta77}. A third rotation
-$\psi_{\rm Eu}$- will align $x_{1,\rm obs}$ and $x_{2,\rm obs}$ with
the symmetry axes of the ellipses projected on the p.o.s. of
the observer (p.o.s.). Eight independent parameters are therefore
required to uniquely geometrically characterize the electron density distribution of
a triaxial galaxy cluster.\\
The axial ratio of the major to the minor axes of the observed
projected isophotes, $e_{\rm proj}(\geq 1)$, is a function of the shape
parameters and of the direction of the l.o.s., defined by the
first two Euler angles, $\theta_{\rm Eu}$ and $\phi_{\rm Eu}$. It is
given by~\citep{Bin80}:
\begin{equation}
\label{eq:tri4e}
e_{\rm proj}= \sqrt{ \frac{j+l + \sqrt{(j-l)^2+4 k^2 } }{j+l
-\sqrt{(j-l)^2+4 k^2 }} },
\end{equation}
where  $j, k$ and $l$ are:
\begin{eqnarray}
j & = &  e_1^2 e_2^2 \sin^2 \theta_{\rm Eu} + e_1^2 \cos^2 \theta_{\rm
Eu} \cos^2 \varphi_{\rm Eu} + e_2^2 \cos^2 \theta_{\rm
Eu} \sin^2 \varphi_{\rm Eu}
\label{eq:tri4a} \\
k & = &  (e_1^2 - e_2^2) \sin \varphi_{\rm Eu} \cos \varphi_{\rm Eu}
\cos \theta_{\rm Eu}
\label{eq:tri4b}  \\
l & = &  e_1^2 \sin^2 \varphi_{\rm Eu} + e_2^2 \cos^2 \varphi_{\rm Eu}
\label{eq:tri4c}
\end{eqnarray}
The rotation angle between the principal axes of the observed ellipses
and the projection onto the sky of the ellipsoid $x_{3,\rm int}$-axis
is~\citep{Bin85}:
\begin{equation}
\label{eq:tri4f}
\psi = \frac{1}{2} \arctan \left[\frac{2 k}{j-l} \right].
\end{equation}
The apparent principal axis that lies furthest from the projection
onto the sky of the $x_{3,\rm int}$-ellipsoid axis is the apparent
major axis if ~\citep{Bin85}
\begin{equation}
(j-l)\cos 2 \psi +2k \sin 2\psi \leq 0
\end{equation}
or the apparent minor axis otherwise. In what follows, we assume
$x_{1,\rm obs}$ to lie along the major axis of the isophotes, so that:
\begin{equation}
\psi_{\rm Eu} = \arctan \left[ \frac{2 k}{j-l-\sqrt{4k^2-(j-l)^2}} \right].
\end{equation}
The projected axial ratio $e_{\rm proj}$, the orientation angle of the
projected ellipses, the slope $\beta$ and the projection in the p.o.s. 
of the core radius can be determined fitting observed
images to the $\beta$-model; further independent constraints are
therefore still required to uniquely determine the gas distribution.\\
The X-Ray surface brightness and the SZ temperature decrement are
given by projection along the l.o.s. of two different powers of
the electron density $n_e$. Following~\cite{Sta77}, we calculate the
projection along the l.o.s. of the electron density distribution,
given by Eq.~(\ref{eq:tri1}), to a generic power $m$ which, in the
observer coordinate system, can be written as:
\begin{equation}
\label{eq:tri5}
\int _{\rm l.o.s.}n_e^m (x_{1,\rm obs},x_{2,\rm obs},l)  dl =  n_{e0}^m \sqrt{\pi}
\frac{\Gamma \left[3 m\beta/2-1/2 \right]}{\Gamma \left[3 m\beta/2
\right]} \frac{D_{\rm c} \theta_{\rm c3}}{\sqrt{h}} \left( 1+ \frac{\theta_{1}^2+e_{\rm proj}^2
\theta_{2}^2}{\theta_{c,\rm proj}^2} \right)^{(1-3 m\beta)/2}
\end{equation}
where $D_{\rm c}$ is the angular diameter distance to the cluster and
$\theta_i \equiv x_{i,\rm obs}/D_{\rm c}$ is the projected angular
position on the p.o.s. of $x_{i,\rm obs}$. $h$ is a function of the
cluster shape and orientation:
\begin{equation}
\label{eq:tri3}
h = e_1^2
\sin^2 \theta_{\rm Eu}
\sin^2
\varphi_{\rm Eu} + e_2^2 \sin^2 \theta_{\rm Eu} \cos^2 \varphi_{\rm Eu} + \cos^2 \theta_{\rm Eu}
\end{equation}
The observed cluster angular core radius $\theta_{c,\rm proj}$ is the
projection on the p.o.s. of the cluster angular intrinsic core radius
\begin{equation}
\label{eq:tri6}
\theta_{\rm c,proj} \equiv \theta_{\rm c3} \left( \frac{e_{\rm proj}}{e_1 e_2} \right)^{1/2}h^{1/4}
\end{equation}
where $\theta_{\rm c3} \equiv r_{\rm c 3}/D_{\rm c}$.

\section{B. Inclination Issues}
\label{sec:inclination}
\begin{figure}
\epsscale{0.5}
\plotone{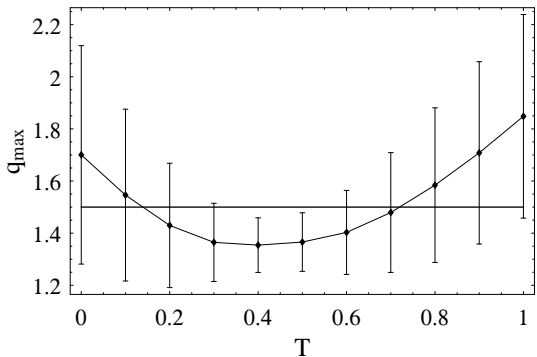}
\caption{The error in the estimate of the axial ratio, $q_\mathrm{max}$, due
to inclination issue vs. the triaxiality degree. The horizontal line
is fixed to the true value of $q_\mathrm{max}$. Points with error bars
refer to estimates in the hypothesis of one principal axis oriented
along the l.o.s.}
\label{fig:f9}
\end{figure}

The analysis presented above is based on the main assumption that one
cluster principal axis is elongated along the line of sight. Despite
this assumption is quite strong, it does not affect significantly the
results. To test the effect of inclination on the estimate of the axis
ratios, we proceed in the following way. First, we generate a galaxy
cluster, characterized by the maximum axis ratio $q_\mathrm{max}$ and
by a parameter related to the degree of triaxiality, $T \equiv \left(
e_\mathrm{mid} - e_\mathrm{min}
\right)/\left( e_\mathrm{max} - e_\mathrm{min} \right)$; oblate
and prolate clusters correspond to $T=0$ and 1, respectively. Since we
are only interested in inclination issues, we neglect other
measurements errors. Then, we generate a set of 25 viewing angles
$\left\{
\theta_\mathrm{Eu}, \varphi_\mathrm{Eu} \right\}$. We assume that
orientations are completely random, i.e. the angle
$\theta_\mathrm{Eu}$ is between 0 and $\pi$ and follows the
distribution $\sin \theta_\mathrm{Eu}/2$, whereas
$\varphi_\mathrm{Eu}$ follows a uniform distribution between 0 and $2
\pi$. For each pair of orientation angles, we compute the projected
ellipticity $e_\mathrm{proj}$ and the elongation $e_\mathrm{l.o.s.}$
and, then, we obtain an estimate of the axial ratios of the cluster in
the hypothesis of one principal axis being aligned along the l.o.s.
Finally, we calculate mean and standard deviation of the set of axial
ratios corresponding to different viewing angles. If the value of such
a mean is near that of the simulated cluster, then inclination issues
hardly affect our analysis. In Fig.~\ref{fig:f9}, we plot the effect on
the estimate of $q_\mathrm{max}$ for different values of T in the case
of $q_\mathrm{max}=1.5$. Error bars equal standard deviations. As we
can see, the error is minimum for highly triaxial clusters ($T \sim
0.5$), being $\sim 0.1$. Such an error should be
added in quadrature to the value of $\Delta q_\mathrm{max}$ estimated
in the previous sections, but due to its smallness it does not
contribute significantly. The error increases for oblate or prolate
clusters, with values $\stackrel{<}{\sim} 0.4$. In this
paper we have been facing with the hypothesis of triaxial clusters, in fact
the algorithm illustrated in the previous sections is optimized to
describe triaxial spheroids. The capability of ellipsoids of revolution to reproduce the
observed data set will be the subject of a forthcoming paper. 
This considerations are quite general and
still holds for very different values of $q_\mathrm{max}$. The error
due to inclination issue on the estimate of axial ratios can be
usually neglected with respect to other uncertainties, in particular
in the measurement of the SZ temperature decrement.

\section{C. Gravitational Lensing}
\label{sec:lensing}
Clusters of galaxies act as lenses deflecting light rays from
background galaxies. In contrast to SZE and X-Ray emission,
gravitational lensing does not probe directly the ICM distribution but
maps the cluster total mass. The ICM distribution in clusters of
galaxies traces the gravitational potential. Since we are considering
a triaxial $\beta$-model for the gas distribution, the gravitational
potential turns out to be constant on a family of similar, concentric,
coaxial ellipsoids. Ellipsoidal potentials are widely used in
gravitational lensing analyses to fit multiple image systems
\citep{sef}.\\
The distribution of the cluster total mass can be inferred from its
gas distribution. If the intra-cluster gas is assumed to be isothermal
and in hydrostatic equilibrium in the cluster gravitational potential,
while non-thermal processes are assumed not to contribute
significantly to the gas pressure, the total dynamical mass density
reads:
\begin{equation}
\label{eq:gl1}
\rho_{\rm tot} = -\left( \frac{k_{B} T_{\rm e}}{4 \pi G \mu m_{\rm p}} \right)
\nabla^2 \left( \ln n_{\rm e} \right)
\end{equation}
where $G$ is the gravitational constant and $\mu m_{\rm p}$ is the
mean particle mass of the gas. If we assume that the electron density
of the ICM follows a $\beta$-model distribution given by
Eq.~(\ref{eq:tri1}), the total gravitating mass density, in the
coordinate system relative to the cluster, is given by:
\begin{equation}
\rho_M = \frac{3 \beta k_{\rm B} T_{\rm e}}{4 \pi G \mu m_{\rm p} r_{\rm c3}^2}
\left(1+\frac{r_{\rm ell}^2}{r_{\rm c3}^2}\right)^{-1} \left[ \sum_{i=1}^3 e_i^2 -\frac{2}{r_{\rm c3}^2}
\frac{\sum_{i=1}^3 \left(
e_i^2 x_{i,\rm int}\right)^2}{1+ r_{\rm ell}^2/r_{\rm c3}^2} \right]
\label{eq:gl2}
\end{equation}
where $e_3=1$ and $r_{\rm ell}$ is the ellipsoidal radius, $r_{\rm
ell}^2 \equiv \sum_{i=1}^3\left( e_i x_{i,\rm int}\right)^2$. The
projected surface mass density can subsequently be written, in the
observer reference frame, as:
\begin{equation}
\label{eq:gl3}
\Sigma = \Sigma_0
\left(1+ \frac{e_{\rm proj}^2}{1+e_{\rm proj}^2}
\frac{\theta_{1}^2+ \theta_{2}^2}{\theta_{\rm c,proj}^2} \right)
\left( 1+
\frac{\theta_{1}^2+e_{\rm proj}^2 \theta_{2}^2}{\theta_{\rm c,proj}^2} \right)^{-3/2}
\end{equation}
where:
\begin{equation}
\label{eq:gl4}
\Sigma_0 = \frac{3}{4} \frac{ \beta k_{\rm B} T_{\rm e}}{G \mu m_{\rm p}}
\frac{\sqrt{ e_1 e_2} }{h^{3/4}}
\frac{1+e_{\rm proj}^2}{\sqrt{e_{\rm proj}}}\frac{1}{\theta_{\rm c,proj}} \frac{1}{D_{\rm c}}
\end{equation}
Although the hypotheses of hydrostatic equilibrium and isothermal gas
are very strong, total mass densities obtained under such assumptions
can yield accurate estimates even in dynamically active clusters with
irregular X-Ray morphologies. Elliptical potentials motivated by X-Ray
observations were employed in the irregular cluster AC~114 to provide
good fit to multiple image systems \citep{DeF04}.\\
The lensing effect is determined by the convergence $k$:
\begin{equation}
k=\frac{\Sigma}{\Sigma_{\rm cr}}
\label{eq:gl4bis}
\end{equation}
which is the cluster surface mass density in units of the surface
critical density $\Sigma_{\rm cr}$:
\begin{equation}
\label{eq:gl5}
\Sigma_{\rm cr} \equiv \frac{c^2}{4 \pi G} \frac{D_{\rm s}}{D_{\rm c} D_{\rm cs}}
\end{equation}
where $D_{\rm cs}$ is the angular diameter distance from the lens to
the source and $D_{\rm s}$ and $D_{\rm c}$ are the angular diameter
distances from the observer to the source and to the lens,
respectively. Fitting the observed surface mass density to a multiple
image system, it is possible to determine the central value of the
convergence:
$$k_0(z_s)=\frac{\Sigma_0}{\Sigma_{\rm cr}}$$
which, using Eqs.~(\ref{eq:gl4}) and~(\ref{eq:gl5}), can be written
as:
\begin{equation}
k_0(z_s)=\frac{ 3 \pi \beta k_{\rm B} T_{\rm e}}{c^2 \mu m_{\rm p}}
\frac{ \sqrt{ e_1 e_2} }{h^{3/4}}
\frac{1+e_{\rm proj}^2}{\sqrt{e_{\rm proj}}} \frac{1}{\theta_{c,\rm proj}}
\frac{D_{\rm cs}}{D_{\rm s}}
\label{eq:gl6b}
\end{equation}



\end{document}